# Near Field Lenses in Two Dimensions

*JB Pendry and SA Ramakrishna*
*The Blackett Laboratory*
*Imperial College*
*London SW7 2BZ*
*UK*

*Abstract*

It has been shown that a slab of materials with refractive index = -1 behaves like a perfect lens focussing all light to an exact electromagnetic copy of an object. The original lens is limited to producing images the same size as the object, but here we generalise the concept so that images can be magnified. For two dimensional systems, over distances much shorter than the free space wavelength, we apply conformal transformations to the original parallel sided slab generating a variety of new lenses. Although the new lenses are not 'perfect' they are able to magnify two dimensional objects. The results apply equally to imaging of electric or magnetic sub wavelength objects in two dimensions. The concepts have potential applications ranging from microwave frequencies to the visible.

1. **Prior Art**

The function of a conventional lens is to collect electromagnetic fields emitted by an object and to refocus them to an image at a new location. See figure 1.

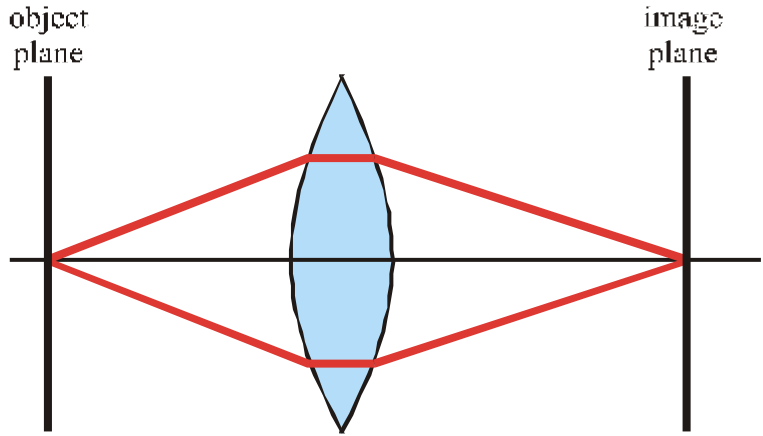

**Figure 1.** A conventional lens refracts beams of light emitted by an object to refocus them in the image plane.

For a conventional lens perfection is defined in terms of: the precision with which the beams are guided to a focus, the aperture which determines how much of the emitted radiation is accepted, and the reduction in spurious reflections from the surface of the lens. We shall assume that we are working at a fixed frequency so that chromatic aberration is not an issue. However even the best made lenses have a limitation: they cannot resolve details in the image an a length scale finer than the wavelength of light at the operating frequency. This comes about because when we solve Maxwell's equations for the wave vector we get,

$$|\mathbf{k}|^2 = k_0^2 = \omega^2 / c_0^2 \qquad (1)$$

If we define the axis of the lens to coincide with the $z$-axis, we can solve for the $z$-component of the wave vector,

$$k_z = \pm\sqrt{k_0^2 - k_x^2 - k_y^2}, \qquad k_0^2 > k_x^2 + k_y^2 \qquad (2)$$

and hence $\theta$, the angle between the ray and the lens axis, is given by,

$$\cos\theta = k_z / k_0 \qquad (3)$$

Lenses with large apertures capture large values of $\theta$ and have better resolution. The maximum value of $\theta$ occurs for $k_z = 0$ at which point the component of the wave vector perpendicular to the lens axis is also a maximum,

$$k_x^2 + k_y^2 = k_0^2, \qquad k_z = 0 \qquad (4)$$

and it is $k_x, k_y$ that define the resolution in the image plane. Hence even for a lens with an infinite aperture there is a natural limit to resolution of,

$$\Delta = \frac{2\pi}{k_{\max}} = \frac{2\pi}{k_0} = \lambda_0 \qquad (5)$$

In the past it has been assumed that this limitation is fundamental, but in a recent paper [1] one of us showed that it is possible to construct a lens that is capable of overcoming the diffraction limit and can in principle approach a perfect electromagnetic reconstruction of the object.

These high resolution details are indeed present in the vicinity of the object, but they correspond to an imaginary wave vector,

$$k_z = \pm i\sqrt{k_x^2 + k_y^2 - k_0^2}, \quad k_0^2 < k_x^2 + k_y^2 \tag{6}$$

which implies that the fields decay exponentially away from the object. These decaying components are usually referred to as the 'near field'. To reconstruct an image including the near field contributions requires an amplifier to restore fields to their original intensity. In fact there is a very simple and elegant device that performs the necessary amplification: take a slab of material in which the electrical permittivity and the magnetic permeability have the following values,

$$\varepsilon = -1, \quad \mu = -1 \tag{7}$$

Some, manipulation of Maxwell's equations will show that such a material also has a refractive index of,

$$n = -1 \tag{8}$$

Veselago [2] was the first to apply Snell's law at the interfaces to show that this slab brings light to a double focus: once inside the slab (assuming that it is thick enough) and once outside. See figure 2.

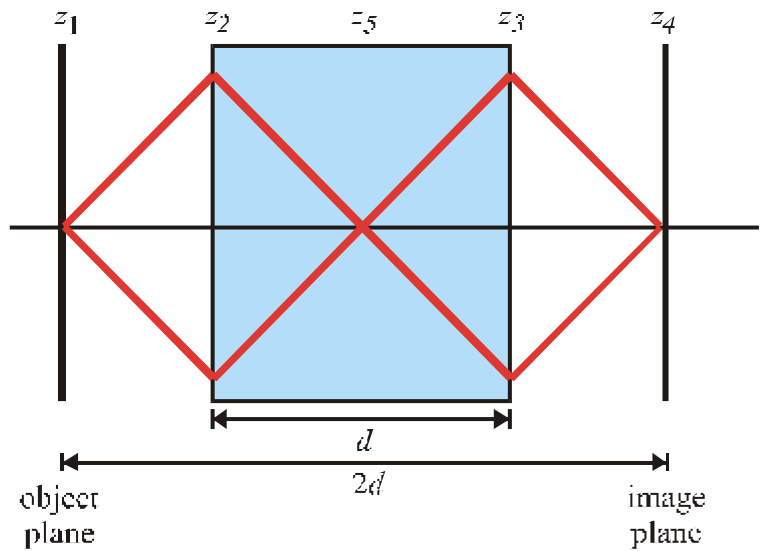

**Figure 2.** A slab of material with refractive index $n = -1$ mimics the action of a conventional lens by refracting beams of light emitted by an object to refocus them in the image plane. For light which is not part of the far field it has highly unconventional behaviour: see figure 3.

The feasibility of making materials with these electromagnetic properties has been demonstrated in the micro wave regime. Materials with negative $\varepsilon$ have been proposed and demonstrated [3], and the same is true for negative $\mu$ [4,5,6,7] and these have been combined to produce negative refraction [8,9].

Ray tracing arguments merely show that the slab does the job of a conventional lens in bringing the rays to a focus. Further investigations show that the slab has another desirable property: there are absolutely no stray reflections at the surfaces so that all light hitting the slab reaches the image. However a lens of a given thickness, $d$, is restricted as to the position of the object and image: no object at a greater distance than $d$ from the nearest surface of the lens can produce an image, and conversely no image can be produced further than $d$ from the nearest surface of the lens.

However the really startling fact about the slab is that the image it produces is perfect [1]. Calculations show that both the near fields and far fields are present in the image. Since the near fields decay exponentially from the object this implies that the slab is behaving as an amplifier! Figure 3 shows a schematic version of the near fields around the slab.

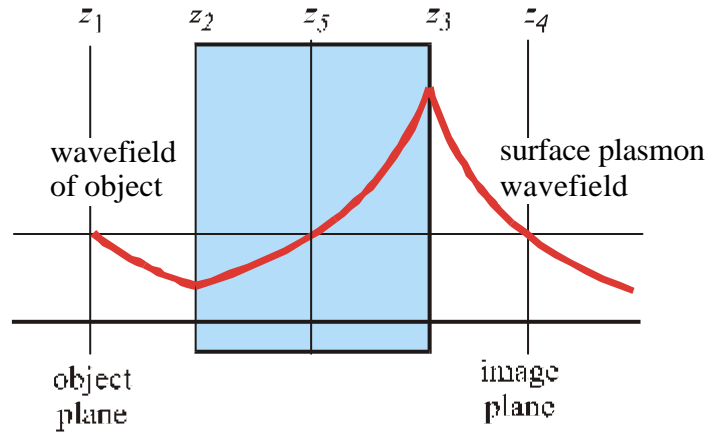

**Figure 3.** For light which is not part of the far field, such as the evanescent fields sketched in the figure, a slab of material with refractive index $n = -1$ amplifies the near fields so that in the image plane they are restored their original values. In this way sub-wavelength details of the object are reconstructed, i.e. refocused, in the image plane. Amplification takes place by stimulating a surface plasmon resonance on the far surface of the slab.

From the figure we can see that amplification occurs through a resonant process: a surface mode [10] is excited on the side of the slab remote from the object and it is this resonance which provides the necessary amplification. Like all resonances this one is frequency specific and therefore the lens is perfect only for the frequency to which the resonance is tuned. In the limit of the extreme near field where,

$$k_x^2 + k_y^2 >> k_0^2 \qquad (9)$$

and for P-polarisation, the resonance corresponds to a surface plasmon mode. The lens is also extremely susceptible to any losses in the system which will damp the resonances. Absorption in effect limits the resolution of any practical realisation of the perfect lens, but still leaves the capability of sub-wavelength imaging. Another limitation of the lens is the planar configuration. This excludes the possibility of magnification which always requires a curved surface. If we try to apply conventional design techniques to generalise the perfect lens we find that the surfaces generated will focus only the far field and the near field is left behind. The reason is that surface plasmon resonances are vital to operation of the lens and curved surfaces in general have a surface plasmon spectrum completely different from that of a planar slab and which is unsuitable for processing the near fields. A different approach is needed.

First we review the perfect lens theory which we generalise to cylindrical surfaces. Then we identify conformal transformations as a tool for generating new lenses, and finally we discuss some realisations and how they can image line charges or line currents.

2. **Focussing the Near Field**

With a conventional lens we can draw a simple picture of rays propagating through the lens and being bent to a focus of the far side, but the perfect lens must also act on the near field and this is a much less familiar entity that cannot be described in terms of ray diagrams. As figure 3 shows the near field reaches the image via a resonant process that has little to do with ray diagrams.

There is another unfamiliar aspect to the near field: whereas in the far field electric and magnetic energy is present in equal measure, in the near field either the electric or magnetic component tends to dominate. In fact in the extreme limit expressed by equation (9) the near fields separate into electrostatic and magnetostatic fields. For example we can write the fields in region 1 between the object and slab (see figure 3) in terms of *S* and *P* polarised contributions:

$$\mathbf{E}(\mathbf{r},t) = \sum_{k_x,k_y} \begin{matrix}[a_S(k_x,k_y)\mathbf{e}_S(k_x,k_y) + a_P(k_x,k_y)\mathbf{e}_P(k_x,k_y)] \\ \times \exp(ik_z z + ik_x x + ik_y y - i\omega t)\end{matrix} \tag{10}$$

$$\mathbf{H}(\mathbf{r},t) = \sum_{k_x,k_y} \left\{ \begin{matrix}[a_S(k_x,k_y)\mathbf{h}_S(k_x,k_y) + a_P(k_x,k_y)\mathbf{h}_P(k_x,k_y)] \\ \times \exp(ik_z z + ik_x x + ik_y y - i\omega t)\end{matrix} \right\} \tag{11}$$

*S* polarisation is defined so that the electric field is normal to the plane of incidence on a surface,

$$\begin{aligned}\mathbf{e}_S^+(k_x,k_y) &= [k_y, \ -k_x, \ 0]\exp\left(ik_x x + ik_y y - \sqrt{k_x^2 + k_y^2 - k_0^2}\,z\right) \\ \mathbf{e}_S^-(k_x,k_y) &= [k_y, \ -k_x, \ 0]\exp\left(ik_x x + ik_y y + \sqrt{k_x^2 + k_y^2 - k_0^2}\,z\right)\end{aligned} \tag{12}$$

and using,

$$\mathbf{k} \times \mathbf{E} = +\omega\mu\mu_0 \mathbf{H} \tag{13}$$

we can deduce the associated magnetic fields,

$$\begin{aligned}\mathbf{h}_S^+(k_x,k_y) &= \frac{1}{\omega\mu\mu_0}\left[-ik_x\sqrt{k_x^2+k_y^2-k_0^2}, \ \left(k_x^2+ik_y\sqrt{k_x^2+k_y^2-k_0^2}\right), \ \left(k_x^2+k_y^2\right)\right] \\ &\quad \exp\left(ik_x x + ik_y y - \sqrt{k_x^2+k_y^2-k_0^2}\,z\right) \\ \mathbf{h}_S^-(k_x,k_y) &= \frac{1}{\omega\mu\mu_0}\left[+ik_x\sqrt{k_x^2+k_y^2-k_0^2}, \ \left(k_x^2-ik_y\sqrt{k_x^2+k_y^2-k_0^2}\right), \ \left(k_x^2+k_y^2\right)\right] \\ &\quad \exp\left(ik_x x + ik_y y + \sqrt{k_x^2+k_y^2-k_0^2}\,z\right)\end{aligned} \tag{14}$$

Under extreme near field conditions, see (9), the magnetic field is the dominant component and the electric field can be neglected.

Conversely for *P* polarisation we define,

$$\begin{aligned}\mathbf{h}_S^+(k_x,k_y) &= [k_y, \ -k_x, \ 0]\exp\left(ik_x x + ik_y y - \sqrt{k_x^2 + k_y^2 - k_0^2}\,z\right) \\ \mathbf{h}_S^-(k_x,k_y) &= [k_y, \ -k_x, \ 0]\exp\left(ik_x x + ik_y y + \sqrt{k_x^2 + k_y^2 - k_0^2}\,z\right)\end{aligned} \tag{15}$$

and use,

$$\mathbf{k} \times \mathbf{H} = -\omega\varepsilon\varepsilon_0 \mathbf{E} \tag{16}$$

to deduce the electric fields,

$$\mathbf{e}_S^+(k_x, k_y) = \frac{-1}{\omega\mu\mu_0}\left[-ik_x\sqrt{k_x^2+k_y^2-k_0^2},\ \left(k_x^2+ik_y\sqrt{k_x^2+k_y^2-k_0^2}\right),\ \left(k_x^2+k_y^2\right)\right]$$

$$\exp\left(ik_x x + ik_y y - \sqrt{k_x^2+k_y^2-k_0^2}\, z\right)$$

$$\mathbf{e}_S^-(k_x, k_y) = \frac{-1}{\omega\mu\mu_0}\left[+ik_x\sqrt{k_x^2+k_y^2-k_0^2},\ \left(k_x^2-ik_y\sqrt{k_x^2+k_y^2-k_0^2}\right),\ \left(k_x^2+k_y^2\right)\right] \quad (17)$$

$$\exp\left(ik_x x + ik_y y + \sqrt{k_x^2+k_y^2-k_0^2}\, z\right)$$

For P polarisation we see that as $\left(k_x^2+k_y^2\right)$ increases and we move to the extreme near field limit, the electric field dominates and the magnetic field becomes irrelevant.

The extreme near field limit is interesting for several reasons. First it contains the very essence of the perfect lens: in this regime the lens focuses details that are inaccessible to a conventional lens. Secondly the dominance of one or other of the fields eases the conditions for perfect focussing: in the case of *P* polarisation magnetic fields are irrelevant in this limit and therefore it is only the electric condition which matters,

$$\varepsilon = -1 \qquad (18)$$

Thirdly in the extreme limit the lens can be generalised in an elegant and simple manner to lift the restrictions of the slab geometry and the rather unhelpful condition that the object and image are of equal size. After all we usually think of a lens as able to magnify an object and the perfect lens does not oblige in this respect.

In this limit, for *P* polarisation, only the electrostatic fields matter which enables us to describe the electrostatic field in terms of a potential. Consider an image which comprises a distribution of electric charge in the plane $z = 0$, which is uniform in *y* but varies with *x*, and which gives rise to an electrostatic field. With no lens in place let the field around the object be,

$$\phi(x,z) = \sum_{k_x} b_1^+(k_x)\exp\left[ik_x x - k_x(z-z_1)\right], \quad 0 < z \qquad (19)$$

where from equations (4), (6), and assuming that $k_y = 0$, we see that,

$$\lim_{k_x \to \infty} k_z = \pm ik_x \qquad (20)$$

and we have chosen the solution satisfying the causal boundary conditions at infinity.

In the presence of a lens,

$$\phi_1(x,z) = \sum_{k_x} b_1^+(k_x)\exp\left[ik_x x - k_x(z-z_1)\right] + b_1^-(k_x)\exp\left[ik_x x - k_x(z-z_1)\right], \quad z_1 < z < z_2,$$

$$\phi_2(x,z) = \sum_{k_x} b_2^+(k_x)\exp\left[ik_x x - k_x(z-z_2)\right] + b_2^-(k_x)\exp\left[ik_x x - k_x(z-z_2)\right], \quad z_2 < z < z_3, \quad (21)$$

$$\phi_3(x,z) = \sum_{k_x} b_3^+(k_x)\exp\left[ik_x x - k_x(z-z_3)\right], \qquad z_3 < z$$

where $z_1, z_2$ mark the beginning and end of the lens. We now match at the boundary to satisfy,

$$\phi_1(x, z_2) = \phi_2(x, z_2), \quad \frac{\partial}{\partial z}\phi_1(x, z_2) = \varepsilon \frac{\partial}{\partial z}\phi_2(x, z_2) = -\frac{\partial}{\partial z}\phi_2(x, z_2),$$
$$\phi_2(x, z_3) = \phi_3(x, z_3), \quad \frac{\partial}{\partial z}\phi_2(x, z_3) = \varepsilon \frac{\partial}{\partial z}\phi_3(x, z_3) = -\frac{\partial}{\partial z}\phi_3(x, z_3)$$
(22)

and obtain,

$$b_3^+(k_x) = \frac{4\varepsilon \exp k_x(-z_1 + z_3)}{(\varepsilon+1)^2 - (\varepsilon-1)^2 \exp k_x(+2z_2 - 2z_3)} b_1^+(k_x)$$

$$b_2^+(k_x) = \frac{2(\varepsilon+1)\exp k_x(-z_2 + z_1)}{(\varepsilon+1)^2 - (\varepsilon-1)^2 \exp k_x(+2z_2 - 2z_3)} b_1^+(k_x)$$

$$b_2^-(k_x) = \frac{2(\varepsilon-1)\exp k_x(+z_1 + z_2 - 2z_3)}{(\varepsilon+1)^2 - (\varepsilon-1)^2 \exp k_x(+2z_2 - 2z_3)} b_1^+(k_x)$$

$$b_1^-(k_x) = \frac{(\varepsilon^2-1)[-\exp(-2k_x z_2) + \exp(-2k_x z_3)]}{(\varepsilon+1)^2 - (\varepsilon-1)^2 \exp k_x(+2z_2 - 2z_3)} b_1^+(k_x)$$
(23)

We note that, in the limit $\varepsilon \to -1$, the wavefields reduce to,

$$\phi_1(x, z) = \sum_{k_x} b_1^+(k_x) \exp[ik_x x - k_x(z - z_1)], \qquad z_1 < z < z_2,$$

$$\phi_2(x, z) = \sum_{k_x} b_1^+(k_x) \exp[+k_x z_1 + ik_x x + k_x(z - z_2)], \qquad z_2 < z < z_3, \quad (24)$$

$$\phi_3(x, z) = \sum_{k_x} b_1^+(k_x) \exp[+k_x(z_1 - z_2 + z_3) + ik_x x - k_x(z - z_3)], \quad z_3 < z$$

and passage through the lens *increases* the amplitude of a given component by a factor of,

$$\exp[+k_x(z_3 - z_2)] \quad (25)$$

and therefore the image is refocused in the plane,

$$z = z_1 + 2(z_3 - z_2) \quad (26)$$

The region between the two images, $z_5 < z < z_4$, is of interest because here the Fourier components are 'over amplified' and in fact with increasing wave vector $k_x$ their amplitude diverges exponentially. These fields do not correctly represent the fields behind the object and are the price we pay for trying to reproduce the fields from a set of charges. There is of course no way that a lens can project a real charge into free space and these divergent fields represent the compromise which has to be made. Much has been made of this point in a recent Physical Review Letter [11]. However when we recognise that real materials will always have a small degree of absorption represented by an imaginary component of,

$$\varepsilon = -1 + i\delta \quad (27)$$

eq (23) tells us that,

$$b_3^+(k_x) \approx \frac{\exp k_x(-z_1 + z_3)}{\frac{1}{4}\delta^2 + \exp[-2k_x(z_3 - z_2)]} b_1^+(k_x) \tag{28}$$

and therefore the amplification process is only valid provided that,

$$\delta^2 << 4\exp[-2k_x(z_3 - z_2)] \tag{29}$$

otherwise the process collapses and for large values of $k_x$ the waves are in fact attenuated. This collapse of amplification prevents the divergence in energy density which would otherwise be implicit in our result. The mathematical divergence is avoided for any finite value of $\delta$. Absorption also limits the spatial resolution obtainable to of order,

$$\Delta = \frac{1}{k_x} = \frac{2d}{-\ln \delta} \tag{30}$$

So for a high resolution lens we must get close to the object (small *d*) and ensure that we use low loss materials.

3. **The Cylindrical Annulus Lens**

At the heart of the operation of the lens is the surface plasmon. In the near field limit the frequencies of all surface plasmons on a planar surface are degenerate,

$$\omega_{sp}(k_x, k_y) = \omega_p/\sqrt{2} \tag{31}$$

where $\omega_p$ is the frequency of the bulk plasmon. The condition that,

$$\varepsilon = 1 - \frac{\omega_p^2}{\omega^2} = -1 \tag{32}$$

ensures that we are tuned to the surface plasmon frequency. There is another circumstance in which all of the surface plasmons are degenerate and this is the surface of a cylinder. Therefore we might ask whether a cylinder has some analogous behaviour to the slab. Once again we assume the near field limit and consider the geometry shown in figure 4.

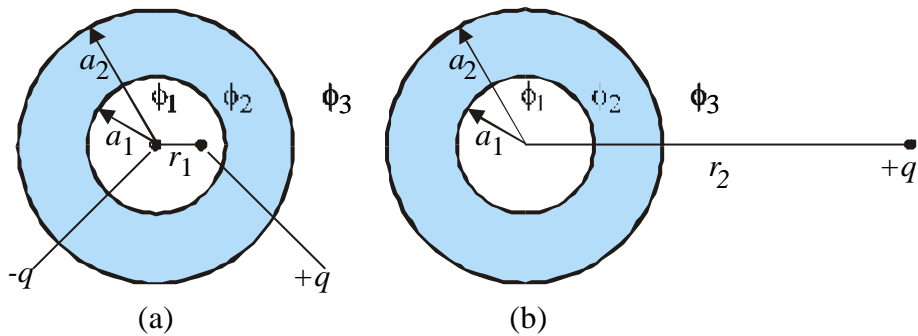

**Figure 4.** A cylindrical annulus of material, dielectric constant $\varepsilon$. (a) the annulus contains two charges, $\pm q$, one charge at the origin, the other distance $r_1$ from the origin. (b) A charge $+q$ sits at a distance $r_2$ from the origin, beyond the annulus.

We shall show that this geometry has a lens like action.

We expand the potential as follows:

$$\phi_1 = \frac{-q}{2\pi\varepsilon_0} \sum_{n=1}^{\infty} \left[ \frac{r_1^n}{nr^n} + c_{1n}^+ r^n \right] \cos n\theta, \qquad r_1 < r < a_1,$$

$$\phi_2 = \frac{-q}{2\pi\varepsilon_0} \sum_{n=1}^{\infty} \left[ c_{2n}^- r^{-n} + c_{2n}^+ r^n \right] \cos n\theta, \qquad a_1 < r < a_2, \qquad (33)$$

$$\phi_3 = \frac{-q}{2\pi\varepsilon_0} \sum_{n=1}^{\infty} c_{3n}^- r^{-n} \cos n\theta, \qquad a_2 < r$$

where the first term in the first equation represents the expansion of the free space potential of a line dipole as shown in figure 4a. Matching the potentials across the two boundaries gives,

$$c_{1n}^+ = \frac{2(\varepsilon^2 - 1) r_1^n (1 - a_1^{-2n})}{n\left((\varepsilon+1)^2 - (\varepsilon-1)^2 a_1^{2n}\right)}, \qquad c_{2n}^+ = \frac{2 r_1^n (\varepsilon - 1)}{n\left[(\varepsilon+1)^2 - (\varepsilon-1)^2 a_1^{2n}\right]},$$

$$c_{2n}^- = \frac{2 r_1^n (\varepsilon + 1)}{n\left[(\varepsilon+1)^2 - (\varepsilon-1)^2 a_1^{2n}\right]}, \qquad c_{3n}^- = \frac{4\varepsilon r_1^n}{n a_2^{-2n} \left((\varepsilon+1)^2 - (\varepsilon-1)^2 a_1^{2n}\right)} \qquad (34)$$

*Limiting case* $\varepsilon \to -1$

$$c_{1n}^+ = 0, \qquad c_{2n}^+ = a_1^{-2n} \frac{r_1^n}{n}, \qquad c_{2n}^- = 0,$$

$$c_{3n}^- = \left[\frac{a_2}{a_1}\right]^{2n} \frac{r_1^n}{n} \qquad (35)$$

which gives,

$$\phi_1 = \frac{-q}{2\pi\varepsilon_0} \sum_{n=1}^{\infty} \frac{r_1^n}{nr^n} \cos n\theta, \qquad r_1 < r < a_1,$$

$$\phi_2 = \frac{-q}{2\pi\varepsilon_0} \sum_{n=1}^{\infty} \frac{r_1^n r^n}{n a_1^{2n}} \cos n\theta, \qquad a_1 < r < a_2, \qquad (36)$$

$$\phi_3 = \frac{-q}{2\pi\varepsilon_0} \sum_{n=1}^{\infty} \left[\frac{a_2}{a_1}\right]^{2n} \frac{r_1^n}{nr^n} \cos n\theta, \qquad a_2 < r$$

i.e. to the external world the system looks like a pair of line charges, one component at the origin, and the second component at a radius,

$$r_2 = r_1 \left[\frac{a_2}{a_1}\right]^2 \qquad (37)$$

In other words the annulus behaves like a lens bringing fields inside the annulus to a focus outside in exactly the same manner that a slab of material produces images of objects on the far side. Furthermore our new lens has the property of magnification by a factor,

$$\left[\frac{a_2}{a_1}\right]^2 \qquad (38)$$

As a further test of the analogy we consider the complementary geometry of a line source outside the annulus as shown in figure 4b. Once more we expand the potential as follows:

$$\phi_1 = \frac{-q}{2\pi\varepsilon_0} \sum_{n=1}^{\infty} c_{1n}^+ r^n \cos n\theta - \frac{q}{2\pi\varepsilon_0} \ln r_2, \qquad r < a_1,$$

$$\phi_2 = \frac{-q}{2\pi\varepsilon_0} \sum_{n=1}^{\infty} \left[ c_{2n}^- r^{-n} + c_{2n}^+ r^n \right] \cos n\theta - \frac{q}{2\pi\varepsilon_0} \ln r_2, \quad a_1 < r < a_2, \qquad (39)$$

$$\phi_3 = \frac{-q}{2\pi\varepsilon_0} \sum_{n=1}^{\infty} \left[ \frac{r^n}{nr_2^n} + c_{3n}^- r^{-n} \right] \cos n\theta - \frac{q}{2\pi\varepsilon_0} \ln r_2, \qquad a_2 < r < r_2$$

where the first term in the last equation represents the expansion of the free space potential of a line charge. Next we must match the potentials across the two boundaries to give,

$$c_{1n}^+ = \frac{2a_2^{2n}}{nr_2^n} \frac{2\varepsilon}{-(1-\varepsilon)^2 a_1^{2n} + (1+\varepsilon)^2 a_2^{2n}}, \qquad c_{2n}^+ = \frac{1}{nr_2^n} \frac{2(1+\varepsilon)a_2^{2n}}{-(1-\varepsilon)^2 a_1^{2n} + (1+\varepsilon)^2 a_2^{2n}},$$

$$c_{2n}^- = \frac{1}{nr_2^n} \frac{2(\varepsilon-1)a_1^{2n} a_2^{2n}}{-(1-\varepsilon)^2 a_1^{2n} + (1+\varepsilon)^2 a_2^{2n}}, \qquad c_{3n}^- = \frac{1}{nr_2^n} \frac{(\varepsilon^2-1)(a_1^{2n} - a_2^{2n})a_2^{2n}}{-(1-\varepsilon)^2 a_1^{2n} + (1+\varepsilon)^2 a_2^{2n}} \qquad (40)$$

*Limiting case* $\varepsilon \to -1$

$$c_{1n}^+ = \frac{1}{nr_2^n} \frac{a_2^{2n}}{a_1^{2n}},$$

$$c_{2n}^+ = 0, \qquad c_{2n}^- = \frac{1}{nr_2^n} a_2^{2n}, \qquad c_{3n}^- = 0 \qquad (41)$$

which gives,

$$\phi_1 = \frac{-q}{2\pi\varepsilon_0} \sum_{n=1}^{\infty} \frac{a_2^{2n}}{a_1^{2n}} \frac{r^n}{nr_2^n} \cos n\theta - \frac{q}{2\pi\varepsilon_0} \ln r_2, \qquad r < a_1,$$

$$\phi_2 = \frac{-q}{2\pi\varepsilon_0} \sum_{n=1}^{\infty} a_2^{2n} \frac{1}{nr_2^n r^n} \cos n\theta - \frac{q}{2\pi\varepsilon_0} \ln r_2, \quad a_1 < r < a_2, \qquad (42)$$

$$\phi_3 = \frac{-q}{2\pi\varepsilon_0} \sum_{n=1}^{\infty} \frac{r^n}{nr_2^n} \cos n\theta - \frac{q}{2\pi\varepsilon_0} \ln r_2, \qquad a_2 < r < r_2$$

i.e. inside the annulus for,

$$r < r_2 \left[ \frac{a_1}{a_2} \right]^2$$

and except for a constant potential shift, the system looks like a line charge located at radius,

$$r_1 = r_2 \left[ \frac{a_1}{a_2} \right]^2 \qquad (43)$$

In other words the annulus behaves like a lens bringing fields outside the annulus to a focus inside and has the property of de-magnification by a factor,

$$\left[\frac{a_1}{a_2}\right]^2 \tag{44}$$

## 4. A New Class of Lenses by Conformal Transformation

The mapping between focussing by a slab and focussing by a cylindrical annulus is so uncannily complete that one suspects a deeper relationship between the two cases, and this is in fact the case. We shall show that the original lens in the form of a slab of $\varepsilon = -1$ dielectric, can be mapped by conformal transformations into a whole class of cylindrical lenses with different focussing properties.

First a refresher course in conformal transformations. A conformal transformation,

(i) preserves the solutions of Laplace's equation in each coordinate system

(ii) Preserves the requirements of continuity of $E_\parallel, D_\perp$ across a boundary with the same dielectric function.

Therefore if we consider an analytic function $\phi(z)$ it will satisfy,

$$\frac{\partial^2 \phi_r}{\partial x^2} + \frac{\partial^2 \phi_r}{\partial y^2} = 0 \tag{45}$$

and if we make a coordinate transformation,

$$z' = u + iv = f(z) \tag{46}$$

it will still be the case that $\phi_r'(z')$ satisfies Laplace's equation in the new coordinate system,

$$\frac{\partial^2 \phi_r'}{\partial u^2} + \frac{\partial^2 \phi_r'}{\partial v^2} = 0 \tag{47}$$

*provided that* the transformation is everywhere analytic in the region under consideration. We shall exploit this invariance under transformation to solve for new geometries.

The other independent component of the near field is the magnetic field. Everything we have said above for imaging a 2D system of electric charges applies to a 2D system of magnetic poles and $\mu = -1$. Of more relevance there is also a mapping between line charges and line currents because not only does the real part of $\ln z$ represent the potential due to a line charge at the origin (or a line magnetic pole) but the imaginary part also represents the vector potential of the magnetic field due to a line current:

$$\mathbf{A}(z) = \begin{bmatrix} 0, & 0, & A_z(z) \end{bmatrix} \tag{48}$$

where,

$$A_z(z) = \mathrm{Im}\ln z = i\theta \tag{49}$$

and hence,

$$\mathbf{B} = \nabla \times \mathbf{A} = \left[\frac{\partial A_z}{\partial y}, \ \frac{-\partial A_z}{\partial x}, \ 0\right] = \left[\frac{-i\cos\theta}{|z|}, \ \frac{+i\sin\theta}{|z|}, \ 0\right] \tag{50}$$

which is the magnetic field due to a line current at the origin. Furthermore in the regions where there is no line current and the permeability is constant,

$$0 \approx \frac{\partial \mathbf{D}}{\partial t} = \nabla \times \mathbf{H} = \mu^{-1}\mu_0^{-1}\nabla \times \mathbf{B} = \mu^{-1}\mu_0^{-1}\left[0, \quad 0, \quad \frac{\partial^2 A_z}{\partial x^2} + \frac{\partial^2 A_z}{\partial y^2}\right] \quad (51)$$

Hence $A_z$ obeys Laplace's equation,

$$\frac{\partial^2 A_z}{\partial x^2} + \frac{\partial^2 A_z}{\partial y^2} = 0 \quad (52)$$

Therefore all the results we derive for imaging line charges apply equally to the magnetic fields created by line currents.

5. **The New Lenses**

*example 1: the cylindrical annulus*

Consider the transformation,

$$z' = \ln z \quad (53)$$

which maps circles centred on the origin in the $z$-plane into lines parallel to the $u$ axis (we must make multiple circuits of the circle to increase the angle without limit). See figures 5(a) and 5(b).

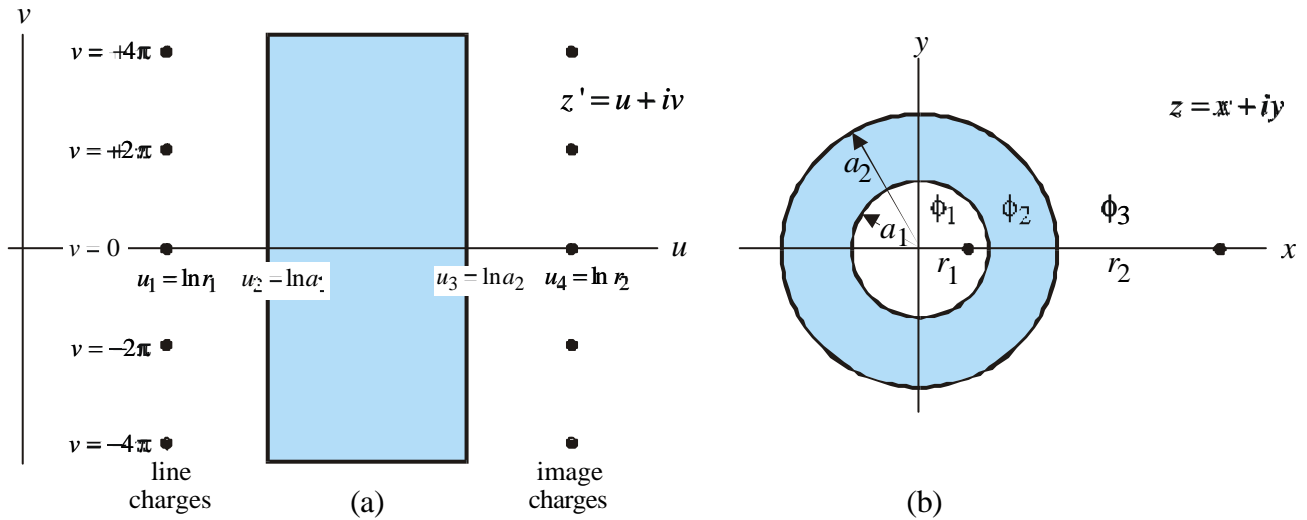

**Figure 5.** (a) In the $uv$ plane we consider a slab of material with dielectric function $\varepsilon = -1$. To the left there is a periodic array of line charges spaced as shown. An image of the line charges is formed to the left of the slab as we have demonstrated above. This configuration can be mapped by conformal transformation (53) into that of (b) producing the configuration of a circular annulus which refocuses the object charge located outside the annulus, to an image located inside the annulus.

From the conformal transformation we deduce that a set of line charges in the $z'$ plane will transform to a single charge of the same magnitude in the $z$ plane but located at distances given by the properties of the transformation,

$$u_1 = \ln r_1, \quad u_2 = \ln a_1, \quad u_3 = \ln a_2, \quad u_4 = \ln r_2 \quad (54)$$

From the focussing properties of the planar lens,

$$u_4 - u_1 = 2(u_3 - u_2) \tag{55}$$

therefore,

$$r_2/r_1 = (a_2/a_1)^2 \tag{56}$$

Therefore a simple conformal transformation generates the results we obtained so laboriously by matching fields at boundaries for the cylindrical case, which is just a mapping of a particular slab geometry.

A line charge in the $z'$ plane is represented by,

$$\phi' = \ln(z' - z'_0) \tag{57}$$

and in the $z$-plane,

$$\phi = \ln(\ln z - z'_0) \tag{58}$$

Expanding about the singularity gives,

$$\phi \approx \ln\{[z - \exp(z'_0)]\exp(-z'_0)\} = \ln[z - \exp(z'_0)] - z'_0 \tag{59}$$

hence the mapping preserves the strength of the singularity and therefore the magnitude and sign of the charge.

*example 2: the crescent lens*

Consider the transformation,

$$z = 1/z' = \frac{u - iv}{u^2 + v^2} \tag{60}$$

which maps the plane $u = a$ into circles:

$$x = \frac{a}{a^2 + v^2}, \quad y = \frac{-v}{a^2 + v^2} \tag{61}$$

hence,

$$\frac{x}{y} = \frac{-v}{a}, \quad v = \frac{-ay}{x} \tag{62}$$

and,

$$x = \frac{a}{a^2 + (ay/x)^2}, \quad ax = \frac{x^2}{x^2 + y^2} \tag{63}$$

or,

$$0 = x^2 + y^2 - \frac{x}{a} = \left(x - \frac{1}{2a}\right)^2 + y^2 - \frac{1}{4a^2} \tag{64}$$

which represents a circle centred on $u = 1/2a$, radius.

Consider a "perfect lens" configuration in the $z'$-plane shown in figure 6(a).

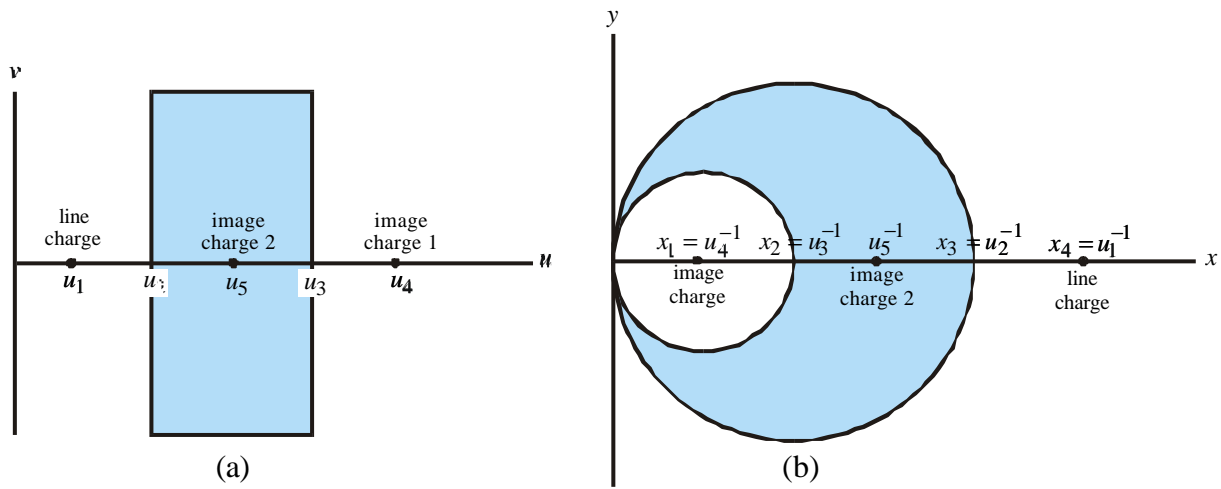

**Figure 6. (a)** The starting configuration of a single line charge and image to which we are going to apply our second transformation to produce the configuration shown in **(b)**.

which will transform in the *z*-plane to a line charge in the *z'* plane is represented by,

$$\phi' = \ln(z' - z'_0) \tag{65}$$

and in the *z*-plane,

$$\phi' = \ln(1/z - z'_0) = \ln\left(-z'_0 \frac{z - z'^{-1}_0}{z}\right) = \ln(-z'_0) + \ln(z - z'^{-1}_0) - \ln z \tag{66}$$

See figure 6(b). In other words unit line charges transform into unit line charges but at shifted positions. There is also a charge at the origin which represents the 'charge at infinity' in the original system. It is removed when the system as a whole is charge neutral. Dipoles, quadrapoles etcetera all change their strength because the notional separation of charges in the multipole is stretched.

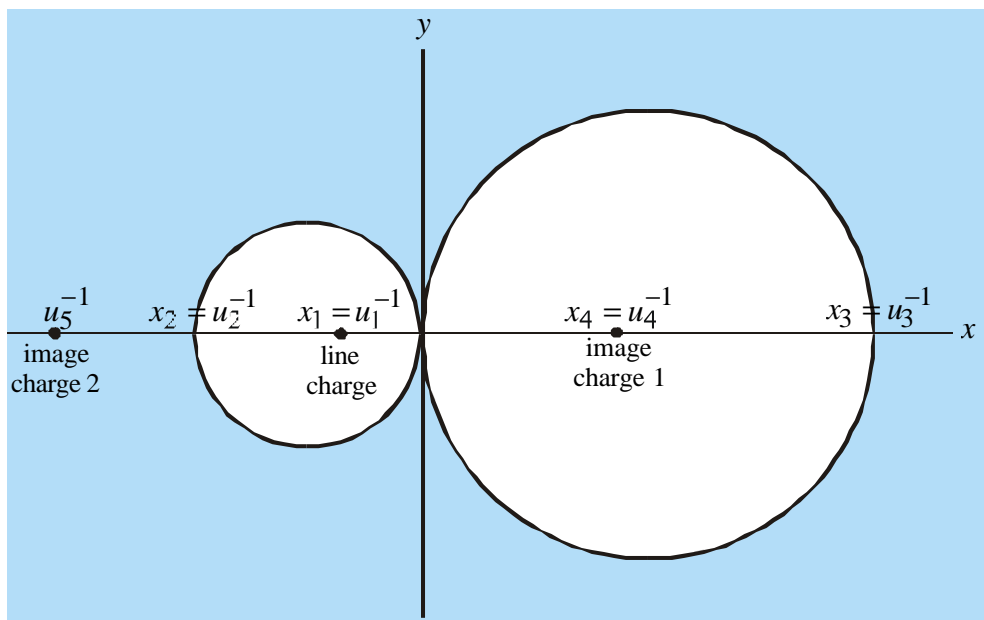

**Figure 7.** A slightly different choice of the parameters in transformation (60) leads to a configuration of two touching cylinders. An object within one cylinder is refocused inside the second cylinder.

*example 3: the kissing lens*

As for the transformation shown in figure 6, but the origin of coordinates is placed inside the slab: see figure 7. This produces a configuration of two touching cylindrical holes embedded in an $\varepsilon = -1$ medium stretching to infinity. This third example takes line charges inside one cylinder and refocuses them inside the second cylinder.

*example 4 - a problem case*

In our enthusiasm for transforming shapes it is easy to run away with the notion that the original slab can be transformed into any 2D shape we desire. This is not the case as the following example makes clear. Consider the transformation,

$$\frac{2}{z'} = \frac{1}{z-b} + \frac{1}{z+b} = \frac{2z}{z^2 - b^2} \tag{67}$$

or,

$$z = \frac{z' \pm \sqrt{z'^2 + 4b^2}}{2} \tag{68}$$

This transformation takes a circle into an ellipse:

$$\begin{aligned} z' = z - b^2 z^{-1} &= r\exp(i\theta) - b^2 r^{-1} \exp(-i\theta) \\ &= \left[r - b^2 r^{-1}\right]\cos(\theta) + i\left[r + b^2 r^{-1}\right]\sin(\theta) \end{aligned} \tag{69}$$

Hence a circle radius $r$ in the $z$-plane is traced in the $z'$-plane as an ellipse. We may seek to apply this transformation to the annulus of figure 5(b), the result appearing in figure 8.

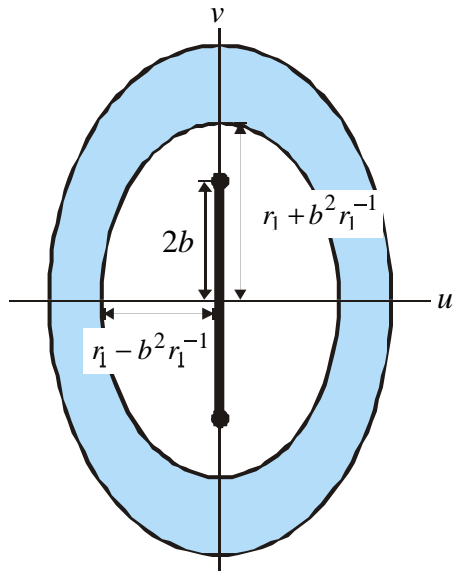

**Figure 8.** The conformal transformation of (67) maps the concentric cylinders of figure 5(b) into ellipses. A problem arises because the transformation is not singled valued but has a cut in the centre of the ellipse across which the potential is in general discontinuous and therefore not a solution of Laplace's equation as required.

The mapping is as follows: all circles in the *z*-plane with $r > b$ map into the entire $z'$-plane, as are circles with $r < b$. The $z'$-plane has a cut from $z' = +2ib$ to $z' = -2ib$.

The doubly connected nature of the problem means that potentials in the *z*-plane that are symmetric about the *x*-axis will in general have a discontinuity in electric field across the cut and therefore are not valid solutions of Laplace's equation in the $z'$-plane.

6. **Using the New Lenses**

Consider the annular lens shown in figure 5. It can be used either to produce an image inside of the lens of an external object, or to produce an image outside an internal object. In the former case the image is demagnified, in the latter case there is magnification. In each case the effect is given by the ratio of the radii of the two cylinders, $a_1^2/a_2^2$; see figure 9(a). Or to put it another way: there is a mapping of the internal space,

$$a_1 > r > a_1^2/a_2 \tag{70}$$

into the external space,

$$a_2 < r < a_2^2/a_1 \tag{71}$$

Apart from the uniform magnification, objects are undistorted by the lens; see figure 9(b). Clearly for given $a_1, a_2$ the lens is limited in its range that is to say myopic.

One obvious application would be to image electrostatic fields on a surface internal to the annulus. This might be a useful trick to enhance the resolution of a scanning near field optical microscope by the magnification of the lens.

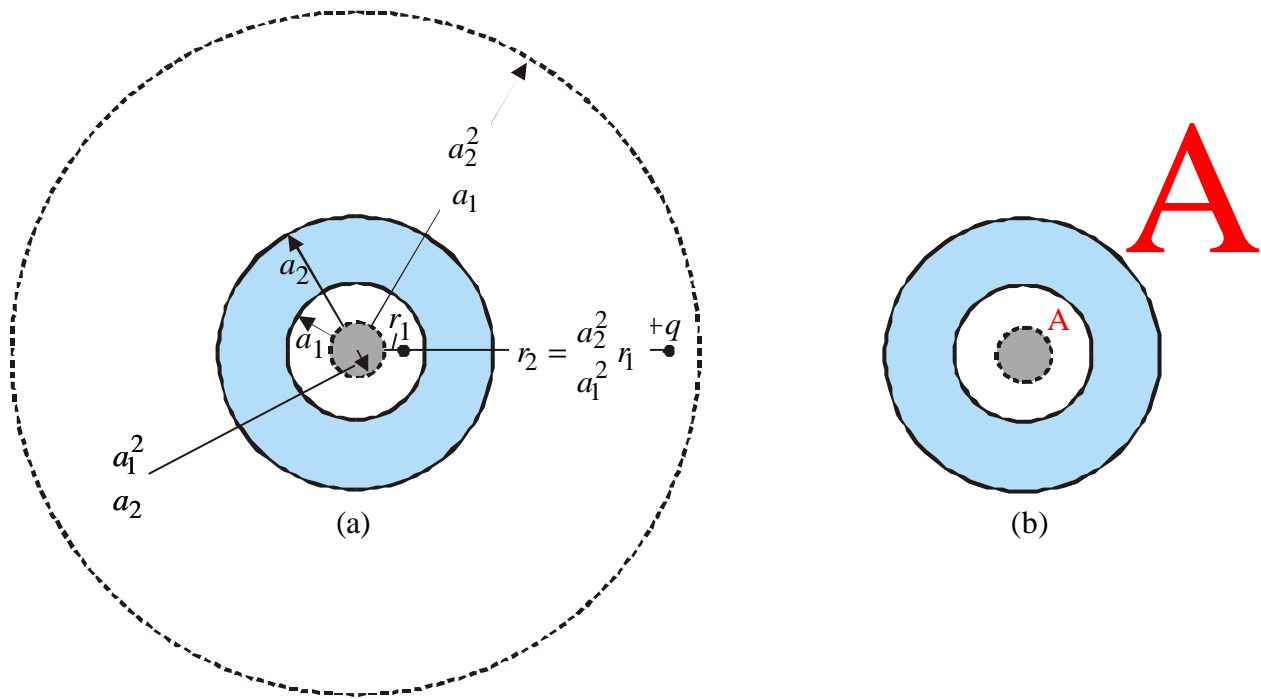

**Figure 9.** (a) The annular lens is myopic: only objects closer than $r = a_2^2/a_1$ can form an image inside the annulus. Conversely objects within the annulus and closer to the centre than $r = a_1^2/a_2$ to the centre will not produce an image outside the annulus. The limiting circles are shown in dashed lines. (b) The annular lens produces a magnified image of internal objects, and a demagnified image of external objects.

Another interesting property of the annular lens is its response to a uniform electric field. From eq (42) we can see that is we have an applied field of,

$$\mathbf{E}_0 = -\nabla \phi_0 = -\nabla \left( E_0 \frac{r}{r_2} \cos\theta \right) \tag{72}$$

then in the limit $\varepsilon \to -1$ the other fields become,

$$\phi_1 = -E_0 \frac{a_2^2}{a_1^2} \frac{r}{r_2} \cos\theta, \qquad r < a_1,$$

$$\phi_2 = -E_0 \frac{a_2^2}{r_2 r} \cos\theta, \qquad a_1 < r < a_2, \tag{73}$$

$$\phi_3 = -E_0 \frac{r}{r_2} \cos\theta, \qquad a_2 < r < r_2$$

In other words the lens is acting as a field amplifier: uniform external fields appear inside the annulus as uniform fields amplified by a factor of $a_2^2/a_1^2$.

The converse of this situation is a line dipole located at the centre of the annulus. reference to equation (36) shows that in the limit $\varepsilon \to -1$ the fields become,

$$\phi_1 = \frac{-q}{2\pi\varepsilon_0} \frac{r_1}{r} \cos\theta, \qquad r_1 < r < a_1,$$

$$\phi_2 = \frac{-q}{2\pi\varepsilon_0} \frac{r_1 r}{a_1^2} \cos\theta, \qquad a_1 < r < a_2, \tag{74}$$

$$\phi_3 = \frac{-q}{2\pi\varepsilon_0} \left[\frac{a_2}{a_1}\right]^2 \frac{r_1}{r} \cos\theta, \qquad a_2 < r$$

and once again we see an amplification by a factor of $a_2^2/a_1^2$ only this time it is the dipole viewed from outside that appears the larger.

Next we examine the crescent lens shown in figure 10. Although its function is similar to that of the annular lens, there are important differences. Firstly the magnification is not uniform but varies according to the position of the object. Therefore in general this lens produces a distorted image. Secondly the regions where imaging can take place have a different geometry. Consideration of the bounds on where the object and image lie for the slab lens and mapping these onto the crescent lens gives for the diameter of the bounding circle,

$$x_{max} = \frac{x_2 x_3}{2x_2 - x_3} \tag{75}$$

and for the inner bound,

$$x_{min} = \frac{x_2 x_3}{2x_3 - x_2} \tag{76}$$

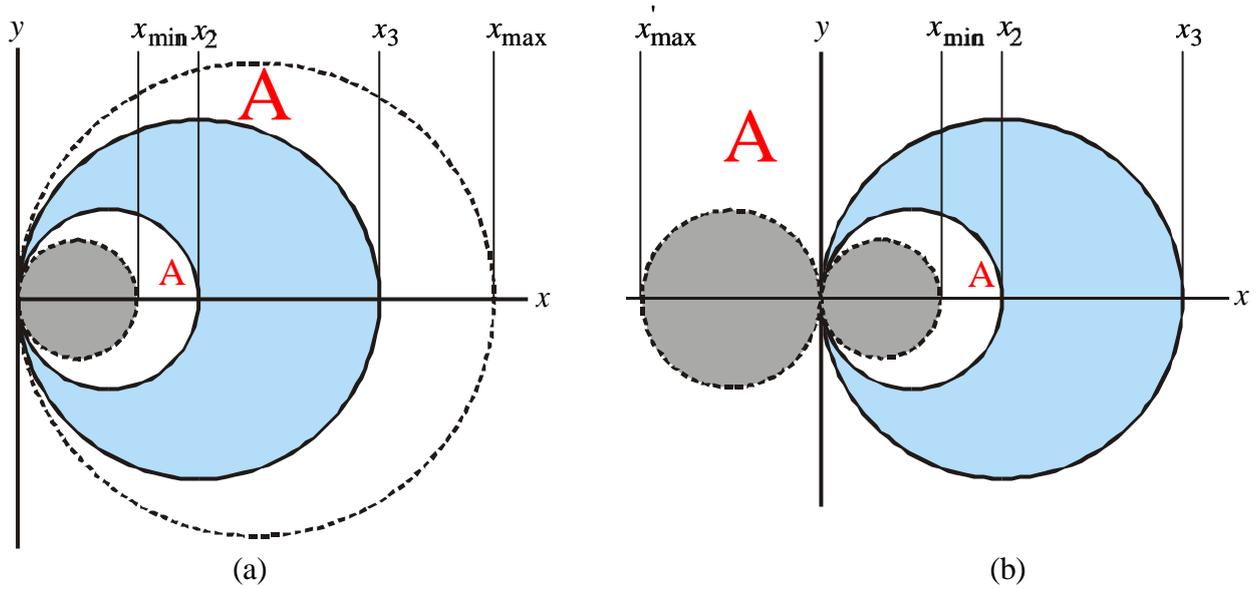

(a)                                                  (b)

**Figure 10.** The crescent lens also has a limit field of view. (a) for $\frac{1}{2}x_3 < x_2 < x_3$, all objects in the crescent between $x_{min}$ and $x_2$ are imaged within the crescent lying between $x_3$ and $x_{max}$ (and vice versa). No other imaging is possible. (b) for $x_1 < \frac{1}{2}x_2$, $x_1 < x_2$ all objects in the crescent between $x_{min}$ and $x_2$ are mapped onto the entire complex plane, excluding the regions inside the circles diameters $x_3$ and $x'_{max}$. A schematic location of object and image is shown in each case.

Hence there are two distinct cases. If the two bounding cylinders of the lens have diameters $x_2, x_3$, see figure 10, and,

$$\tfrac{1}{2}x_3 < x_2 < x_3 \tag{77}$$

then, only objects within a crescent bounded by cylinders diameter $x_{min}$, $x_2$ can produce an image, which then appears in the crescent bounded by cylinders diameter $x_3$, $x_{max}$ (and vice versa). If the inequality (77) is not obeyed so that,

$$x_2 < \tfrac{1}{2}x_3, \quad x_2 < x_3 \tag{78}$$

then, only objects within a crescent bounded by cylinders diameter $x_{min}$, $x_2$ can produce an image, but now the image appears exterior to the area of the complex plane bounded by the two cylinders diameters $x'_{max}, x_3$ (and vice versa). In this instance it is in principle possible for an object to project an image to infinity, or to refocus an object at infinity within the inner crescent. Since an object at $z_{ob}$ is refocused at,

$$z_{im} = \frac{1}{z_{ob}^{-1} - 2\left(x_2^{-1} - x_3^{-1}\right)} \tag{79}$$

it is clear that the vicinity of the point,

$$z_{ob} = \frac{x_2 x_3}{2(x_3 - x_2)} \tag{80}$$

maps to infinity. Obviously the projection of an image of a charge to infinity would be a very singular process and in any case the near field approximation would break down at some distance comparable with the free space wavelength. In practice the limit on how far away we can project an image will depend on the amount of absorption present.

7. **Conclusions**

In the limit of all lengths much shorter than the wave length (the near field limit) we have given recipes for a series of two dimensional lenses derived by conformal transformations of the original slab lens. Just as a conventional lens works only for the far field, so these new lenses specialise to the near field as if the requirement of magnification requires some compromise on the functionality of the lens. Both electric and magnetic versions of the lens can be envisaged the former requiring $\varepsilon = -1$, the latter $\mu = -1$. The lenses have the property of magnifying the image relative to the object and therefore can perform tasks not accessible to the original perfect lens.

We are sure that our list of conformal transformations has not exhausted the interesting possibilities. In particular it might be imagined that a series of lenses could be employed in more sophisticated processes than those considered here.

**Acknowledgements**

We acknowledge support from DoD/ONR MURI grant N00014-01-1-0803.

**References**

[1]  Veselago, VG 1968 Soviet Physics USPEKHI, **10**, 509.

[2]  Pendry JB, 2000 *Phys. Rev. Lett.* **85** 3966.

[3]  Sievenpiper DF, Sickmiller ME, and Yablonovitch E, Phys Rev Lett, 1996 **76,** 2480; Pendry JB, Holden AJ, Stewart WJ, Youngs I, 1996 *Phys Rev Lett* **76** 4773; Pendry JB, Holden AJ, Robbins DJ, and Stewart WJ, 1998 J. Phys. [Condensed Matter], **10,** 4785.

[4]  Pendry JB, Holden AJ, Robbins DJ, and Stewart WJ, 1999 IEEE Transactions on Microwave Theory and Techniques, **47**, 2075.

[5]  Sievenpiper DF, Zhang L, Broas RFJ, Alexopoulos FJ, Yablonovitch E, 1999 IEEE Trans. Micr. Theory and Tech. **47**, 2059-2074.

[6]  Broas RFJ, Sievenpiper DF, Yablonovitch E 2001 IEEE Trans. Micr. Theory and Tech. **49**, 1262-1265.

[7]  Wiltshire MCK, Pendry JB, Young IR, Larkman DJ, Gilderdaleand Hajnal JV., 2001 *Science* **291** 848-51.

[8]  Smith DR, Padilla Willie J, Vier DC, Nemat-Nasser SC, Schultz S, 2000 Phys. Rev. Lett. **84**, 4184.

[9]  Shelby RA, Smith DR, Schultz S, 2001 *Science*, **292**, 79.

[10]  Ritchie RH, 1957 Phys. Rev., **106,** 874.

[11]  Garcia N and Nieto-Vesperinas M, 2002 *Phys. Rev. Lett.*, **88**, 207403.